\documentclass[12pt]{article}

\usepackage{graphicx}

\begin{document}

\begin{center}
{\bf Rational nonlinear electrodynamics causes the inflation of the universe  } \\
\vspace{5mm} S. I. Kruglov
\footnote{E-mail: serguei.krouglov@utoronto.ca}

\vspace{3mm}
\textit{Department of Physics, University of Toronto, \\60 St. Georges St.,
Toronto, ON M5S 1A7, Canada\\
Department of Chemical and Physical Sciences, University of Toronto Mississauga,\\
3359 Mississauga Road North, Mississauga, ON L5L 1C6, Canada} \\
\vspace{5mm}
\end{center}

\begin{abstract}
The source of the universe inflation is electromagnetic fields obeying rational nonlinear electrodynamics proposed earlier. Within this model the singularities of the electric field at the center of charges, the Ricci scalar, the Ricci tensor squared and the Kretschmann scalar are absent. We consider the universe which is filled by stochastic magnetic fields. It is demonstrated that the inflation lasts approximately $2$ s with the reasonable e-folding number $N\approx 63$. The Inflation starts from de Sitter spacetime and after the universe inflation end it decelerates approaching the radiation era.
\end{abstract}

\section{Introduction}

The universe inflation can be explained by modifying general relativity (GR) \cite{Starobinsky} or by introducing a scalar field which drives the inflation \cite{Linde}. Another scenario of the inflation is to modify electrodynamics when electromagnetic fields are very strong in the early time of the universe evolution  \cite{Camara}-
 \cite{Kruglov}. The usage of nonlinear electromagnetic fields can remove the singularities during Big Bang.

This paper is a continuation of the work \cite{Kruglov1} where the universe inflation takes place for the stochastic magnetic background obeying rational nonlinear electrodynamics proposed in \cite{Kruglov5}.

In a relativistic electron-positron plasma there are the stochastic fluctuations of the electromagnetic field, and as a result, plasma fluctuations may be the source of a stochastic magnetic field  \cite{Lemoine}, \cite{Lemoine1}. Thermal fluctuations in the pre-recombination plasma lead to a primordial magnetic field. We suppose that the early universe was filled by a strong random magnetic field in the early stage of the radiation-dominated era. Observations show the existence of magnetic fields in the universe from $10^{-5}$ ps to $10^4$ Mps \cite{Neronov}, \cite{Taylor}. We suppose that the primordial magnetic fields are seeds for the large magnetic fields observed today \cite{Kronberg}, \cite{Lemoine1}. The large magnetic fields have the primordial origin and are generated by the galactic dynamo theory. Thus, it is implied that on large scales primordial magnetic fields are produced in the early universe \cite{Subramanian}. It should be noted that the origin of cosmic magnetism on the largest scales of galaxies, galaxy clusters and the general inter galactic medium is still an open problem \cite{Gaensler}.
The electric field is screened by the charged primordial plasma and, therefore, we consider the case $E = 0$ \cite{Lemoine1}. According to the standard cosmological model the universe is isotropic in the large scale, and therefore we use the equality $\langle\textbf{B}\rangle = 0$. As a result, there are not directional effects.

The structure of the paper is as follows. In Sect. 2 we introduce rational nonlinear electrodynamics (RNED) and estimate the parameter $\beta$ of the model. General relativity coupled to RNED was considered in Sect. 3. Evolution of the universe within our model was studied in Sect. 4. Section 5 is a conclusion.

\section{A model of RNED}

We consider GR coupled with RNED with the Lagrangian \cite{Kruglov5}
\begin{equation}
{\cal L} = -\frac{{\cal F}}{2\beta{\cal F}+1},
\label{1}
\end{equation}
where ${\cal     F}=(1/4)F_{\mu\nu}F^{\mu\nu}=(\textbf{B}^2-\textbf{E}^2)/2$ is a field invariant.
The model of RNED possesses attractive features such as simplicity and non-singularity. Thus, RNED coupled to GR results in the existence of regular magnetic black holes \cite{Kr}. In addition, the size of the shadow of M87* black hole estimated on the base of RNED-GR gives the result which is in agrement with the data of Event Horizon Telescope collaboration \cite{Kr1}.

The energy density $\rho$ and the pressure $p$ for a magnetic background ($E=0$) are given by
\begin{equation}
\rho=\frac{B^2}{2(1+\beta B^2)},~~~p=\frac{B^2(1-3\beta B^2)}{6(1+\beta B^2)^2}.
\label{2}
\end{equation}
We suppose that our model at weak electromagnetic fields is converted into QED with loop corrections. We obtain the model parameters $\beta$ by comparing (1), at the weak field limit, with the Heisenberg$-$Euler Lagrangian. If $\beta{\cal F}\ll 1$ Lagrangian (1) becomes
\begin{equation}
{\cal L}=-{\cal F}+2\beta{\cal F}^2-6\beta^2{\cal F}^3+{\cal O}\left((\beta{\cal F})^4\right).
\label{3}
\end{equation}
The QED Lagrangian with one loop correction (the Heisenberg$-$Euler Lagrangian) is given by \cite{Gies}
\begin{equation}
{\cal L}_{HE}=-{\cal F}+c{\cal F}^2,~~~c=\frac{8\alpha^2}{45m_e^4},
\label{4}
\end{equation}
where $\alpha=e^2/(4\pi)\approx 1/137$ and the electron mass $m_e=0.51~\mbox{MeV}$. By comparing (3) and (4) we find
\begin{equation}
\beta=\frac{4\alpha^2}{45m_e^4}=69\times 10^{-5}~\mbox{MeV}^{-4}.
\label{5}
\end{equation}

\section{Cosmology}

The line element of the homogeneous and isotropic Friedmann$-$Robertson$-$ Walker (FRW) spacetime is given by
\begin{equation}
ds^2=-dt^2+a(t)^2\left(dx^2+dy^2+dz^2\right),
\label{6}
\end{equation}
where $a(t)$ is a scale factor. We assume that in the early stage of the universe the cosmic background is stochastic magnetic fields. The averaged over a volume magnetic fields obey equations
\begin{equation}
\langle\textbf{B}\rangle=0,~~~~\langle E_iB_j\rangle=0,~~~~\langle B_iB_j\rangle=\frac{1}{3}B^2g_{ij}.
\label{7}
\end{equation}
Then nonlinear electrodynamics can be represented as a perfect fluid \cite{Novello1}. It is worth noting that Eq. (7) leads to isotropy because the directionality is absent.
For simplicity we omit the brackets $\langle \rangle$ in the following. The Friedmann equation is given by
\begin{equation}
3\frac{\ddot{a}}{a}=-\frac{\kappa^2}{2}\left(\rho+3p\right).
\label{8}
\end{equation}
When $\rho + 3p < 0$ the universe accelerates. Making use of Eq. (2) we obtain
\begin{equation}
\rho+3p=\frac{B^2(1-\beta B^2)}{(1+\beta B^2)^2}.
\label{9}
\end{equation}
The requirement of the universe acceleration $\rho + 3p < 0$ leads to $\beta B^2>1$. From Eq. (5) we obtain $1/\sqrt{\beta}\approx 6\times 10^{11}$ T. For the strong magnetic fields $B>6\times 10^{11}$ T the universe inflation takes place. The conservation of the energy-momentum tensor, $\nabla^\mu T_{\mu\nu}=0$, leads to the equation
\begin{equation}
\dot{\rho}+3\frac{\dot{a}}{a}\left(\rho+p\right)=0.
\label{10}
\end{equation}
After the integration of Eq. (10), taking into account Eq. (2), we obtain
\begin{equation}
B(t)=\frac{B_0}{a^2(t)},
\label{11}
\end{equation}
with $B_0$ being the magnetic field corresponding to the value $a(t)=1$.
The scale factor increases during the inflation but the magnetic field decreases.
Making use of Eqs. (2) and (11) we find
$
\lim_{a(t)\rightarrow 0}\rho(t)=\lim_{a(t)\rightarrow 0}p(t)=\lim_{a(t)\rightarrow \infty}\rho(t)=\lim_{a(t)\rightarrow \infty}p(t)=0.
$
There are no singularities of the energy density and pressure as $a(t)\rightarrow 0$ and $a(t)\rightarrow \infty$.
With the help of Eq. (2) one obtains the equation of state (EoS) $w=p/\rho$ and its limits
\begin{equation}
w=\frac{1-3\beta B^2}{3(1+\beta B^2)},~~~
\lim_{a\rightarrow\infty} w=\frac{1}{3},~~\lim_{a\rightarrow 0} w=-1.
\label{12}
\end{equation}
We have the EoS for ultra-relativistic case as $a(t)\rightarrow \infty$. As $a(t)\rightarrow 0$  EoS corresponds to de Sitter spacetime $w=-1$. As a result, the inflation starts from  de Sitter spacetime.
 From the Einstein equation one obtains the curvature
\begin{equation}
R=\kappa^2{\cal T}=\kappa^2\left[\rho(t)-3p(t)\right],
\label{13}
\end{equation}
and Eqs. (2) and (13) lead to
$
\lim_{a(t)\rightarrow 0}R(t)=\lim_{a(t)\rightarrow \infty}R(t)=0.
$
As a result, there are not a singularity of the Ricci scalar. The Ricci tensor squared $R_{\mu\nu}R^{\mu\nu}$ and the Kretschmann scalar $R_{\mu\nu\alpha\beta}R^{\mu\nu\alpha\beta}$ are expressed by linear combinations of $\kappa^4\rho^2$, $\kappa^4\rho p$, and $\kappa^4p^2$ \cite{Kruglov1} and they vanish as $a(t)\rightarrow 0$ and $a(t)\rightarrow \infty$. Making use of Eqs. (9) and (11) we make a conclusion that the universe inflation occurs at $a(t)<\beta^{1/4}\sqrt{B_0}$.

\section{Evolution of the universe}

For three dimensional flat universe the second Friedmann equation is given by
\begin{equation}
\left(\frac{\dot{a}}{a}\right)^2=\frac{\kappa^2\rho}{3}.
\label{14}
\end{equation}
Making use of Eqs. (2) and (14) with the unitless variable $x=a/(\beta^{1/4}\sqrt{B_0})$, Eq. (14) reads
\begin{equation}
\dot{x} =\frac{\kappa x}{\sqrt{6\beta(x^4+1)}}.
\label{15}
\end{equation}
The plot of the function $y\equiv\sqrt{6\beta}\dot{x}/\kappa$ versus $x$ is depicted in Fig. 1.
\begin{figure}[h]
\includegraphics[height=4.0in,width=4.0in]{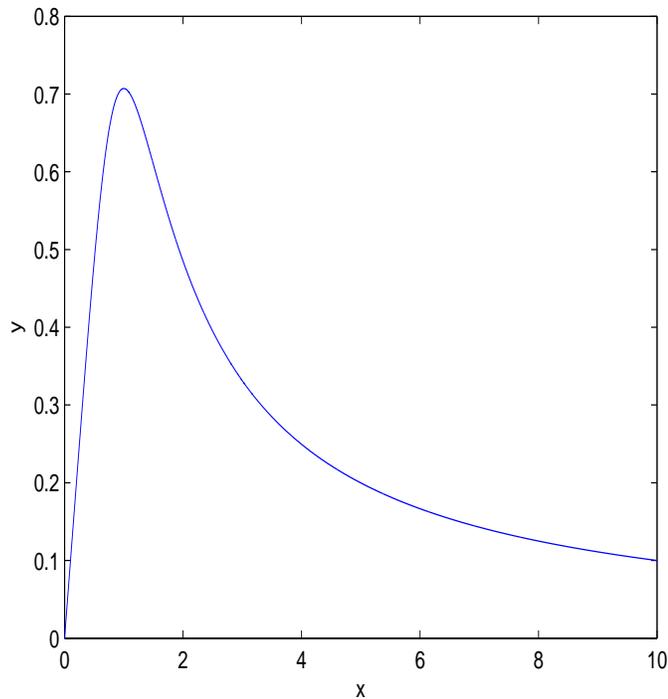}
\caption{\label{fig.1}The function $y\equiv\sqrt{6\beta}\dot{x}/\kappa$ vs. $x$.}
\end{figure}
According to Fig. 1, after Big Bang the universe accelerates ($\dot{y}>0$) until the graceful exit point $x=1$ ($a=(\beta^{1/4}\sqrt{B_0})$) and then the universe decelerates.
After integration of Eq. (15) we find
\begin{equation}
\int_\epsilon^x \frac{\sqrt{(x^4+1)}}{x}dx =\frac{\kappa}{\sqrt{6\beta}}\int_0^t dt,
\label{16}
\end{equation}
where $\epsilon$ corresponds to the starting point of the universe inflation ($t=0$) and $x=1$ (at the time $t$) corresponds to the end of the inflation. We use Eq. (16) to study the evolution of the universe inflation.
It is convenient to consider the deceleration parameter $q$ to study the expansion of the universe. With the help of Eqs. (2), (8),  and (14) we obtain
\begin{equation}
q=-\frac{\ddot{a}a}{(\dot{a})^2}=\frac{x^4-1}{x^4+1}.
\label{17}
\end{equation}The plot of the function (17) is depicted at Fig. 2.
\begin{figure}[h]
\includegraphics[height=4.0in,width=4.0in]{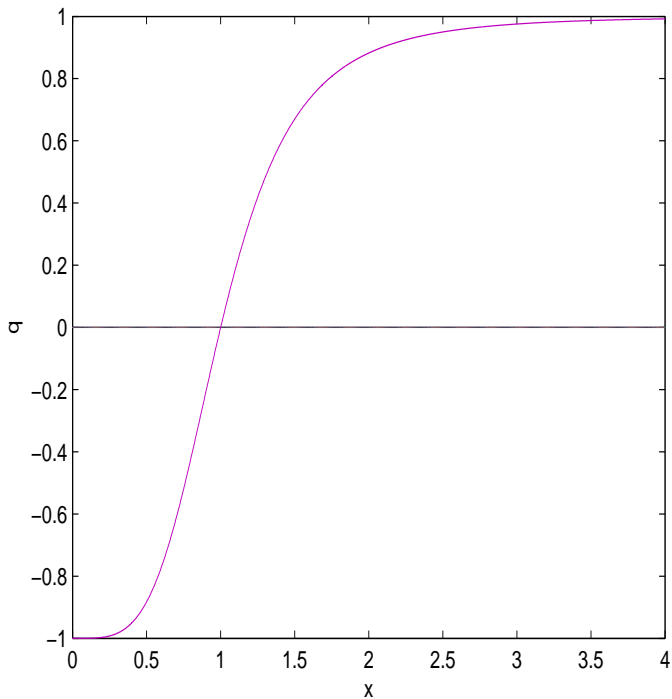}
\caption{\label{fig.2}The function $q$ vs. $x=a/(\beta B_0^2)^{1/4}$.}
\end{figure}
In accordance with Fig. 2 the inflation ($q<0$ ) lasts until the graceful exit $x=1$ and after
 the universe decelerates ($q>0$).

We will estimate the amount of the inflation by using the e-folding number \cite{Liddle}
\begin{equation}
N=\ln\frac{a(t_{end})}{a(t_{in})},
\label{18}
\end{equation}
where $t_{in}$ is an initial time of the inflation and $t_{end}$ is the final time. The graceful exit point is $x=1$ corresponding to $a(t_{end}) b$ ($b\equiv \beta^{1/4}\sqrt{B_0}$).
If e-folding number $N\approx 65$ \cite{Liddle} the horizon and flatness problems are solved.
We will analyze Eq. (16) to estimate the duration of the inflationary period. Using units ($c=\hbar=1$) $\kappa=\sqrt{8\pi G}=4.1\times 10^{-28}~\mbox{eV}^{-1}$, $\beta=6.9\times 10^{-29}~\mbox{eV}^{-4}$ (see Eq. (5)), $1~\mbox{s}=1.5\times 10^{15}~\mbox{eV}^{-1}$, we find $\kappa/\sqrt{6\beta}=0.2\times10^{-13}\mbox{eV}\approx 30~\mbox{s}^{-1}$.
To have the reasonable amount of inflation we use the value $\varepsilon=6.5\times 10^{-28}$ corresponding to the beginning of the inflation. Then from Eq. (16) one finds that the duration of the universe inflation is approximately $2$ s and  the e-folding number $N\approx 63$. This model describes phases of the universe acceleration, deceleration and the graceful exit.
Making use of Eq. (2) we obtain the equation for the perfect fluid
\begin{equation}
p=-\rho+f(\rho),~~~f(\rho)=\frac{4}{3}\rho(1-2\beta\rho).
\label{19}
\end{equation}
When $|f(\rho)/\rho|\ll 1$  during the inflation, the spectral index $n_s$, the tensor-to-scalar ratio $r$, and the running of the spectral index $\alpha_s=dn_s/d\ln k$ have reasonable values \cite{Odintsov} according to the PLANCK experiment \cite{Ade} and WMAP data \cite{Komatsu}, \cite{Hinshaw}.
The inequality $|f(\rho)/\rho|\ll 1$ gives $\beta B^2\gg 1/3$  that corresponds to the inflation phase ($\beta B^2>1$, $B>6\times 10^{11}$ T).

It is worth noting that the general form of the equation for the perfect fluid in the cosmological context was discussed in \cite{Odintsov}, \cite{Odintsov1}, \cite{Ganiou}. One can postulate EoS to describe the inflation similar to one under consideration. But it is important to understand what kind of matter (dark energy) drives the universe to accelerate. In the present paper we specify the matter field to be the RNED which leads to concrete form (19) and leads to universe inflation.

\section{Conclusion}

We demonstrated that our model with homogeneous and isotropic cosmology describes the reasonable amount of the universe inflation ($\approx 2$ s) and e-folding number ($N\approx 63$). There are no singularities of the energy density, pressure, the Ricci scalar, the Ricci tensor squared, and the Kretschmann scalar. A stochastic magnetic field is the source of the universe inflation at the early epoch. The magnetic field decreases as $B=B_0/a(t)^2$  during inflation till the graceful exit and then the universe decelerates approaching to the radiation era.


\begin{thebibliography}{99}

\bibitem{Starobinsky} Alexei A. Starobinsky, Phys. Lett. B  \textbf{91} 99-102 (1980).
\bibitem{Linde} Andrei D. Linde, Lect. Notes Phys. \textbf{738}, 1-54 (2008).

\bibitem{Camara} C. S. Camara, M. R. de Garcia Maia, J. C. Carvalho, and J. A. S. Lima, Phys. Rev. D \textbf{69}, 123504 (2004).


\bibitem{Novello3} V. A. De Lorenci, R. Klippert, M. Novello, and J. M. Salim, Phys. Rev. D \textbf{65}, 063501 (2002).

\bibitem{Novello} M. Novello, S. E. Perez Bergliaffa, and J. M. Salim, Phys. Rev. D \textbf{69}, 127301 (2004).

\bibitem{Novello1} M. Novello, E. Goulart, J. M. Salim, and S. E. Perez Bergliaffa, Class. Quant. Grav. \textbf{24}, 3021 (2007).

\bibitem{Vollick} D. N. Vollick, Phys. Rev. D \textbf{78}, 063524 (2008).

\bibitem{Salcedo} R. Garc\'{i}a-Salcedo, T. Gonzalez, A. Horta-Rangel, and I. Quiros. Phys. Rev. D \textbf{90}, 128301 (2014).

\bibitem{Kruglov1} S. I. Kruglov, Phys. Rev. D \textbf{92}, 123523 (2015).
\bibitem{Kruglov2} S. I. Kruglov, Int. J. Mod. Phys. A \textbf{31}, 1650058 (2016).
\bibitem{Kruglov3} S. I. Kruglov, Int. J. Mod. Phys. D \textbf{25}, 1640002 (2016).
\bibitem{Kruglov4} S. I. Kruglov, Int. J. Mod. Phys. A \textbf{32 }, 1750071 (2017).
\bibitem{Kruglov} S. I. Kruglov, Eur. Phys. J. Plus \textbf{135}, 370 (2020).
\bibitem{Kruglov5} S. I. Kruglov, Ann. Phys. \textbf{353}, 299 (2014).
\bibitem{Lemoine} D. Lemoine, Phys. Rev. D \textbf{51}, 2677 (1995).
\bibitem{Lemoine1} D. Lemoine and M. Lemoine, Phys. Rev. D \textbf{52}, 1955 (1995).
\bibitem{Kronberg} P. P. Kronberg, Rept. Prog. Phys. \textbf{57}, 325 (1994).
\bibitem{Neronov} A. Neronov and I. Vovk, Science \textbf{328}, 73 (2010).
\bibitem{Taylor} A. M. Taylor, I. Vovk, and A. Neronov, Astron. Astrophys. \textbf{529}, A144 (2011).
\bibitem{Subramanian} K. Subramanian, Rept. Prog. Phys. \textbf{79}, 076901 (2016).
\bibitem{Gaensler} B. M. Gaensler, R. Beck, and L. Feretti, New Astronomy Reviews \textbf{48}, 1003 (2004).
\bibitem{Kr} S. I. Kruglov. Dyonic and magnetic black holes with rational nonlinear electrodynamics.     \textit{Preprints} \textbf{2020}, 2020010049 (doi:10.20944/preprints202001.0049.v1).
\bibitem{Kr1} S. I. Kruglov, Mod. Phys. Lett. A (in press); The shadow of M87* black hole within rational nonlinear electrodynamics. \textit{Preprints} \textbf{2020}, 2020040228 (doi:10.20944/preprints202004.0228.v2).
\bibitem{Gies} W. Ditrich and H. Gies, \textit{Probing the Quantum Vacuum} (Springer Tracts in Modern Physics, \textbf{166}, 2000).

\bibitem{Liddle} A. R. Liddle and. H. Lyth, \textit{Cosmological Inflation and Large-Scale Structure} (Cambrige University Press, 2000).

\bibitem{Odintsov} K. Bamba, S. Nojiri, and S. D. Odintsov, Phys. Rev. D \textbf{90}, 124061 (2014).

\bibitem{Ade} P. A. R. Ade et al. (Planck Collaboration), Astron. Astrophys. \textbf{571}, A16 (2014) [arXiv:1303.5083, arXiv:1303.5082, arXiv:1303.5076].

\bibitem{Komatsu} E. Komatsu et al. (WMAP Collaboration), Astrophys. J. Suppl. \textbf{192}, 18 (2011) [arXiv:1001.4538].

\bibitem{Hinshaw} G. Hinshaw et al. (WMAP Collaboration), Astrophys. J. Suppl. \textbf{208}, 19 (2013) [arXiv:1212.5226].

\bibitem{Odintsov} S. Nojiri,, S. D. Odintsov, Phys. Rev.D \textbf{70}, 103522 (2004).
\bibitem{Odintsov1} A. V. Astashenok, S. Nojiri, S. D. Odintsov, R. J. Scherrer, Phys. Lett. B \textbf{713}, 145 (2012).
\bibitem{Ganiou}M. G. Ganiou, M. J. S. Houndjo, I. G. Salako, M. E. Rodrigues, J. Tossa,
Astrophys. Space Sci. \textbf{361},  210 (2016).



\end{thebibliography}
\end{document}